\documentclass[12pt]{article}

\usepackage{amsmath}
\usepackage{stmaryrd}
\usepackage{subfigure}
\usepackage{empheq}
\usepackage{appendix,epic,eepic,amsmath,amsthm,amssymb,epsfig,cite,indentfirst,oldgerm}
\renewcommand{\theequation}{\arabic{equation}}
\newcommand{\EQ}{\begin{equation}}
\newcommand{\EN}{\end{equation}}

\newcommand{\IM}{\mathbf{\imath}}

\newcommand{\bear}{\begin{eqnarray}}
\newcommand{\ear}{\end{eqnarray}}
\newcommand{\bt} { \begin{tabular} }
\newcommand{\et}{ \end{tabular} }
\newcommand{\bc} { \begin{center} }
\newcommand{\ec}{ \end{center} }

\newcommand{\btb} { \begin{table} }
\newcommand{\etb}{ \end{table} }

\begin{document}

\topmargin 0pt
\oddsidemargin 5mm
\newcommand{\NP}[1]{Nucl.\ Phys.\ {\bf #1}}
\newcommand{\PL}[1]{Phys.\ Lett.\ {\bf #1}}
\newcommand{\NC}[1]{Nuovo Cimento {\bf #1}}
\newcommand{\CMP}[1]{Comm.\ Math.\ Phys.\ {\bf #1}}
\newcommand{\PR}[1]{Phys.\ Rev.\ {\bf #1}}
\newcommand{\PRL}[1]{Phys.\ Rev.\ Lett.\ {\bf #1}}
\newcommand{\MPL}[1]{Mod.\ Phys.\ Lett.\ {\bf #1}}
\newcommand{\JETP}[1]{Sov.\ Phys.\ JETP {\bf #1}}
\newcommand{\TMP}[1]{Teor.\ Mat.\ Fiz.\ {\bf #1}}

\renewcommand{\thefootnote}{\fnsymbol{footnote}}

\newpage
\setcounter{page}{0}
\begin{titlepage}
\begin{flushright}

\end{flushright}
\vspace{0.5cm}
\begin{center}
{\large Algebraic Geometry methods associated to the one-dimensional Hubbard model} \\
\vspace{1cm}
{\large M.J. Martins } \\
\vspace{0.15cm}
{\em Universidade Federal de S\~ao Carlos\\
Departamento de F\'{\i}sica \\
C.P. 676, 13565-905, S\~ao Carlos (SP), Brazil\\}
\vspace{0.35cm}
\end{center}
\vspace{0.5cm}

\begin{abstract}
In this paper we study the covering vertex 
model of the one-dimensional Hubbard Hamiltonian
constructed by Shastry in the realm of algebraic geometry. We show that the
Lax operator sits in a genus one curve which is not isomorphic but 
only isogenous to the curve suitable for the AdS/CFT context. We provide an
uniformization of the Lax operator in terms of ratios of
theta functions allowing us to establish
relativistic like properties such as 
crossing and unitarity. We show that the 
respective $\mathrm{R}$-matrix
weights lie on an Abelian surface being birational to the product of
two elliptic curves with distinct $\mathrm{J}$-invariants. 
One of the curves is isomorphic to that
of the Lax operator but the other is solely fourfold isogenous.
These results clarify the reason
the $\mathrm{R}$-matrix can not be written using only difference
of spectral parameters of the Lax operator.

\end{abstract}

\vspace{.15cm} \centerline{}
\vspace{.1cm} \centerline{Keywords: Hubbard Model , Algebraic Geometry}
\vspace{.15cm} \centerline{February 2016}

\end{titlepage}


\pagestyle{empty}

\newpage

\pagestyle{plain}
\pagenumbering{arabic}

\renewcommand{\thefootnote}{\arabic{footnote}}
\newtheorem{proposition}{Proposition}
\newtheorem{pr}{Proposition}
\newtheorem{remark}{Remark}
\newtheorem{re}{Remark}
\newtheorem{theorem}{Theorem}
\newtheorem{theo}{Theorem}

\def\ll{\left\lgroup}
\def\rr{\right\rgroup}

\newtheorem{Theorem}{Theorem}[section]
\newtheorem{Corollary}[Theorem]{Corollary}
\newtheorem{Proposition}[Theorem]{Proposition}
\newtheorem{Conjecture}[Theorem]{Conjecture}
\newtheorem{Lemma}[Theorem]{Lemma}
\newtheorem{Example}[Theorem]{Example}
\newtheorem{Note}[Theorem]{Note}
\newtheorem{Definition}[Theorem]{Definition}

\section{Introduction}

The Hubbard model originates from the tight-binding
formulation for solids where the electrons can hope between
lattice sites but also interact through
the Coulomb repulsion. In its simplest
form, electron hopping takes place between nearest neighbour
sites with the same kinetic energy
while the Coulomb interaction occurs only for electrons at
the same site with a constant strength $\mathrm{U}$.
The Hubbard Hamiltonian on a ring of size $\mathrm{N}$ with interaction 
symmetric under electron-hole 
transformation is given by,
\begin{equation}
\label{HAMH}
\mathrm{H}=-\sum_{j=1}^{\mathrm{N}} \sum_{\sigma=\uparrow,\downarrow} 
(c^{\dagger}_{j\sigma}c_{j+1\sigma} + c^{\dagger}_{j+1\sigma}c_{j\sigma}) 
+\mathrm{U} \sum_{j=1}^{\mathrm{N}}(c^{\dagger}_{j\uparrow}c_{j\uparrow}-\frac{1}{2})  
(c^{\dagger}_{j\downarrow}c_{j\downarrow}-\frac{1}{2}),  
\end{equation}
where $c^{\dagger}_{j\sigma}$ and $c_{j\sigma}$ stand for 
creation and annihilation
operators for an electron at site $j$ with spin $\sigma$. 

In a groundbreaking work Lieb and Wu showed that 
Hamiltonian (\ref{HAMH})
is exactly
diagonalized by means of an extention of the 
coordinate Bethe ansatz method besides the model absence 
of Mott transition \cite{LW}. Over the years this solution
has been used to compute many other physical properties 
and for a recent extensive review we refer to the monograph \cite{EKG}.
Exact integrability from the viewpoint of the quantum 
inverse scattering
approach was only established many years 
later by Shastry
in three influential 
papers \cite{SHA,SHA1,SHA2}. An important result was the 
discovery of a classical two-dimensional
vertex model on the square $\mathrm{N} \times \mathrm{N}$
lattice whose row-to-row transfer matrix commutes 
with the spin version
of the Hubbard Hamiltonian. This spin 
model was
obtained by applying a generalized version
of the Jordan-Wigner transformation on the bulk
term of Eq.(\ref{HAMH}) which can 
be rewritten as \cite{SHA},
\begin{equation}
\label{HAMS}
\mathrm{H} =
\sum_{j=1}^{\mathrm{N}} \sigma^{+}_{j}\sigma^{-}_{j+1} 
+ \sigma^{-}_{j}\sigma^{+}_{j+1} 
+ \tau^{+}_{j}\tau^{-}_{j+1} + \tau^{-}_{j}\tau^{+}_{j+1} + 
\frac{\mathrm{U}}{4}\sigma^{z}_{j}\tau^{z}_{j}, 
\end{equation}
where $\sigma^{\pm}_{j},\sigma^{z}_{j}$ and 
$\tau^{\pm}_{j},\tau^{z}_{i}$ 
are two commuting sets of Pauli matrices 
acting on the site $j$. 
Recall that strict periodic boundary conditions for
electron Hamiltonian (\ref{HAMH}) leads to sector dependent
twisted boundary conditions for 
the spin operator (\ref{HAMS}) and the precise 
form of this relationship
can for instance be found in \cite{MR}.
However, this difference on boundaries can 
be easily captured by 
introducing fermionic statistics into the 
integrable structures without affecting the
main features of Shastry's construction \cite{WA}.

The appealing form of the spin Hamiltonian (\ref{HAMS})
led Shastry
to propose that the underlying classical vertex model 
should be given
by coupling appropriately two six-vertex models
obeying the so-called free-fermion condition. Let us denote by
$\mathrm{L}_{0j}(\omega)$ the Lax operator encoding the  
Boltzmann weights structure of such coupled six-vertex models. 
As usual the indices $0$ and $j$ refer to operators acting on
the auxiliary and
quantum spaces associated respectively with the degrees of freedom
sited on the  horizontal and vertical edges of the square lattice.
In terms of Pauli matrices such Lax operator can be expressed by,
\begin{equation}
\label{LAX1}
\mathrm{L}_{0j}(\omega)=\exp\left[\frac{h}{2}(\sigma_0^{z} \tau_0^{z}+\mathrm{I}_{0})\right] \mathrm{I}_j
\left[\mathcal{L}^{(\sigma)}_{0j}(a,b,c) \mathcal{L}^{(\tau)}_{0j}(a,b,c)\right]
\exp\left[\frac{h}{2}(\sigma_0^{z} \tau_0^{z}+\mathrm{I}_{0})\right] \mathrm{I}_j,
\end{equation}
where $\mathrm{I}$ is the four-dimensional 
identity matrix and 
the symbol $\omega$
denotes the set of parameters $a,b,c$ and $h$.

The Lax operators $\mathcal{L}^{(\sigma)}_{0j}(a,b,c)$  
and $\mathcal{L}^{(\tau)}_{0j}(a,b,c)$ represent the weights of two 
copies of six-vertex models whose expressions are,
\begin{equation}
{\mathcal L}^{\sigma}_{0j}(a,b,c) =
\frac{(a+b)}{2}\mathrm{I}_0\mathrm{I}_j + 
\frac{(a-b)}{2} \sigma^{z}_{0}\sigma^{z}_{j} +
c(\sigma^{+}_{0}\sigma^{-}_{j} + \sigma^{-}_{0}\sigma^{+}_{j}),
\end{equation}
and
\begin{equation}
{\mathcal L}^{\tau}_{0j}(a,b,c) =
\frac{(a+b)}{2}\mathrm{I}_0\mathrm{I}_j + 
\frac{(a-b)}{2} \tau^{z}_{0}\tau^{z}_{j} +
c(\tau^{+}_{0}\tau^{-}_{j} + \tau^{-}_{0}\tau^{+}_{j}),
\end{equation}
such that the so-called free-fermion condition is fullfiled,
\begin{equation}
\label{LAX2}
a^2+b^2=c^2.
\end{equation}

In order to assure integrability the six-vertex 
free-fermion weights $a,b,c$ and the dimensionless
interaction $h$ must be constrained by, 
\begin{equation}
\label{LAX3}
\sinh(2h)=\frac{\mathrm{U}ab}{2c^2}.
\end{equation}

In addition to that, Shastry considered the 
local condition
that is sufficient for the commutativity of two 
transfer matrices built out of Lax operators
with distinct weights parameters.
In fact, the explicit form of 
the $\mathrm{R}$-matrix $\mathrm{R}(\omega_1,\omega_2)$
operator satisfying the Yang-Baxter relation,
\begin{equation}
\label{YBAX}
\mathrm{R}_{12}(\omega_1,\omega_2) \mathrm{L}_{13}(\omega_1) \mathrm{L}_{23}(\omega_2)
=\mathrm{L}_{23}(\omega_2) \mathrm{L}_{13}(\omega_1) \mathrm{R}_{12}(\omega_1,\omega_2),
\end{equation}
has been determined in references \cite{SHA1,SHA2}. 

In recent years new insights into the Hubbard
model emerged from the investigation by Beisert
of integrable structures associated to
the fundamental representation
of centrally extended 
$\mathrm{su}(2|2)$ superalgebra \cite{BEI}. This	
representation depends on the central elements values
which have been parametrized in terms of two
variables $x_{+}$ and $x_{-}$ constrained by
the genus one curve \cite{BEI},
\begin{equation}
\label{CURVE1}
\mathrm{E}_1 \equiv x_{+}+\frac{1}{x_{+}}-x_{-}-\frac{1}{x_{-}}-\IM \mathrm{U}=0.
\end{equation}

Afterwards it has been pointed out that 
the intertwining operator based on such 
representation 
of the $\mathrm{su}(2|2)$ superalgebra 
can be related to the
original Shastry
$\mathrm{R}$-matrix \cite{BEI1}. 
This equivalence was further elaborated in \cite{MM} for
a factorizable $\mathrm{S}$-matrix derived in the context of
the $\mathrm{su}(2|2)$ Zamolodchikov-Faddeev algebra \cite{FRO}.
Such relationship occurs up to gauge transformation 
and when the $\mathrm{R}$-matrices
parameters are identified as \cite{MM},
\begin{equation}
\label{SPEC}
x_{+}= \frac{\IM a \exp(2h)}{b},~~~
x_{-}= \frac{-\IM b \exp(2h)}{a}.
\end{equation}

At this point we recall that this mapping
goes back at least to the parameterization used 
in \cite{SHA2,MR} for the eigenvalues 
of the transfer matrix based on the Lax
operator (\ref{LAX1}-\ref{LAX3}). We also note
that the expression for $\mathrm{E}_1$ is
exactly Eq.(31) of
ref.\cite{MR} taken into account
identification (\ref{SPEC}). 

Although the above connection suggests that the 
Lax operator (\ref{LAX1}-\ref{LAX3}) could
be sited on an elliptic curve it does not mean
that such underlying spectral curve is 
necessarily isomorphic
to $\mathrm{E}_1$. In this paper we shall show
that the right hand side of 
Eq.(\ref{SPEC}) involves quadratic powers on
the polynomial ring variables in which the
Lax operator (\ref{LAX1}-\ref{LAX3}) 
is properly defined. This fact excludes 
isomorphism but leaves the
possibility that the Hubbard model 
spectral curve $\mathrm{E}_2$
be isogenous 
to $\mathrm{E}_1$.
Recall that a $\mathrm{n}$-fold isogeny among elliptic 
curves $\mathrm{E}_2$ and $\mathrm{E}_1$ 
is a surjective morphism that maps the
distinguished point
of $\mathrm{E}_2$ (place at ``infinity")
to the distinguished point of $\mathrm{E}_1$ \cite{SILV}. The integer $\mathrm{n}$ is the degree
of the morphism and thus a generic point of $\mathrm{E}_1$ is mapped to $\mathrm{n}$ distinct
points of $\mathrm{E}_2$.
In fact, it turns out that
the spectral curve underlying 
the Shastry Lax operator is given 
by the following affine quartic elliptic curve,
\begin{equation}
\label{CURVE2}
\mathrm{E}_2 \equiv (x^2+y^2)^2-\mathrm{U}xy-1=0,
\end{equation}
where the suitable ring variables $x$ and $y$ are
related to the weights used by Shastry as,
\begin{equation}
\label{SPEC1}
x=a\exp(h),~~~y=b\exp(h).
\end{equation}

In next section we discuss the derivation of the 
curve $\mathrm{E}_2$ from
the original construction by Shastry 
of the covering Hubbard model. 
We also show that the curves $\mathrm{E}_2$ and $\mathrm{E}_1$ are
not isomorphic but only have a fourth degree isogeny. 
In section \ref{SEC2} we argue that the uniformization
of $\mathrm{E}_2$ can be performed along the lines
of the symmetrical eight vertex model with weights
satisfying the free-fermion condition \cite{BAX}.
The matrix elements of the Lax operator are then represented
in terms of factorized ratios of theta functions. This allows us
to present local inversion properties for the Lax operator such
as crossing and unitarity relations. In section \ref{SEC3}
we discuss the geometrical properties associated with 
the $\mathrm{R}$-matrix
of the Hubbard model. We show that the 
$\mathrm{R}$-matrix weights
lies on an Abelian surface built out of the product of two
non-isomorphic elliptic curves. Our concluding remarks are in section \ref{SEC4} and in
two appendices we present
technical details of some computations omitted in the main text.

\section{Lax operator spectral curve}

The problem of finding 
integrable systems leads us to solve 
a set of polynomial 
relations on the product
of three projective spaces originated from the Yang-Baxter equation. This means that all the
matrix elements of a given Lax operator are expected to be
determined by homogeneous polynomials in suitable ring variables up to
an overall normalization.
Inspecting the entries 
of the Lax operators (\ref{LAX1}-\ref{LAX3})
one concludes that the respective polynomial ring 
is $\mathbb{C}[x,y,c]$ where
the variables $x$ and $y$ have already been 
defined in Eq.(\ref{SPEC1}). Upon this identification
the explicit matrix form of the Lax operator is,
\begin{equation}
\label{LAXO}
{\scriptsize
\mathrm{L}_{12}(x,y,c)=\left(
\begin{array}{cccccccccccccccc}
x^2& 0& 0& 0& 0& 0& 0& 0& 0& 0& 0& 0& 0& 0& 0& 0 \\
0& xy& 0& 0& xc& 0& 0& 0& 0& 0& 0& 0& 0& 0& 0& 0 \\
0& 0& xy& 0& 0& 0& 0& 0& xc& 0& 0& 0& 0& 0& 0& 0 \\
0& 0& 0& y^2& 0& 0& yc& 0& 0& yc& 0& 0& \theta(x,y)& 0& 
0& 0 \\
0& xc& 0& 0& \frac{xyc^2}{\theta(x,y)}& 0& 0& 0& 0& 
      0& 0& 0& 0& 0& 0& 0 \\
0& 0& 0& 0& 0& \frac{x^2c^2}{\theta(x,y)}& 0& 
      0& 0& 0& 0& 0& 0& 0& 0& 0 \\
0& 0& 0& yc& 0& 0& 
      \frac{y^2c^2}{\theta(x,y)}& 0& 0& c^2& 0& 0& yc& 0& 0& 0 \\
     0& 0& 0& 0& 0& 0& 0& \frac{xyc^2}{\theta(x,y)}& 0& 0& 0& 0& 0& 
      xc& 0& 0 \\
0& 0& xc& 0& 0& 0& 0& 0& 
      \frac{xyc^2}{\theta(x,y)}& 0& 0& 0& 0& 0& 0& 0 \\
     0& 0& 0& yc& 0& 0& c^2& 0& 0& \frac{y^2c^2}{\theta(x,y)}& 0& 
      0& yc& 0& 0& 0 \\
0& 0& 0& 0& 0& 0& 0& 0& 0& 0& 
      \frac{x^2c^2}{\theta(x,y)}& 0& 0& 0& 0& 0 \\ 
     0& 0& 0& 0& 0& 0& 0& 0& 0& 0& 0& \frac{xyc^2}{\theta(x,y)}& 0& 0& 
      xc& 0 \\
0& 0& 0& \theta(x,y)& 0& 0& yc& 0& 0& yc& 0& 
      0& y^2& 0& 0& 0 \\
0& 0& 0& 0& 0& 0& 0& xc& 0& 0& 0& 0& 0& 
      xy& 0& 0 \\
0& 0& 0& 0& 0& 0& 0& 0& 0& 0& 0& xc& 0& 0& 
      xy& 0 \\
0& 0& 0& 0& 0& 0& 0& 0& 0& 0& 0& 0& 0& 0& 0& x^2 \\
\end{array}
\right),
}
\end{equation}
where $\theta(x,y)=x^2+y^2$.

The next step is the determination 
of the spectral curve which 
should constrain the variables
$x,y$ and $c$. This task can be 
done by eliminating 
the unwanted variables
$a,b$ and $\exp(2h)$ with the help
Eqs.(\ref{LAX2},\ref{SPEC1}),
\begin{equation}
a=\frac{x}{\exp(h)},~~  
b=\frac{y}{\exp(h)},~~\exp(2h)=\frac{x^2+y^2}{c^2}.  
\end{equation}

By substituting the above results
in Eq.(\ref{LAX3}) we find that the
desired spectral curve is,
\begin{equation}
\label{CURVEP2}
\overline{\mathrm{E}}_2 \equiv (x^2+y^2)^2-\mathrm{U}xyc^2-c^4=0,
\end{equation}
which is just the projective 
closure of the affine curve (\ref{CURVE2}).

Let us now show that the curve
$\overline{\mathrm{E}}_2$ is connected with the 
projective closure of $\mathrm{E}_1$ by means of a fourfold isogeny.
We first note that from Eq.(\ref{CURVE1}) 
the expression for
$\overline{\mathrm{E}}_1$ is given by,
\begin{equation}
\overline{\mathrm{E}}_1=(x_{+}-x_{-})(x_{+}x_{-}-z^2)-\IM \mathrm{U} x_{+} x_{-} z,
\end{equation}
where the variable $z$ refers to the extra projective coordinate.

By using Eqs.(\ref{LAX2},\ref{SPEC},\ref{SPEC1}) we can 
establish the following
morphism between the elliptic curves $\overline{\mathrm{E}}_2$ and
$\overline{\mathrm{E}}_1$,
\EQ
\label{map}
\renewcommand{\arraystretch}{1.5}
\begin{array}{ccc}
\overline{\mathrm{E}}_2 \subset \mathbb{CP}^2[x,y,c] &~~~ \overset{\psi}{\longrightarrow}~~~ 
& \overline{\mathrm{E}}_1 \subset \mathbb{CP}^2[x_{+},x_{-},z] \\
(x:y:c)
& \longmapsto &~(\psi_{1}:\psi_{2}:\psi_{3}),
\end{array}
\EN
where the polynomials map expressions are,
\begin{equation}
\psi_1=\IM x^2(x^2+y^2),~~\psi_2=-\IM y^2 (x^2+y^2),~~\psi_3=xyc^2.
\end{equation}

Note that the above map is defined everywhere even at the singular
points $(1,\pm \IM,0) \in \overline{\mathrm{E}}_2$. In fact, at these 
particular points 
one can find
an alternative representation 
of $\psi$ with the help of the 
polynomial (\ref{CURVEP2}), namely
\begin{equation}
(\psi_1:\psi_2:\psi_3) \sim (x^2:-y^2:-\frac{\IM xyc^2}{x^2+y^2}) 
\sim (x^2:-y^2:-\frac{\IM xy(x^2+y^2)}{c^2+\mathrm{U}xy}), 
\end{equation}
and as result we obtain 
$\psi(0:\pm \IM:0)=(1:1:0) \in \overline{\mathrm{E}}_1$.

The degree of the morphism (\ref{map}) can be determined as
the cardinality of the fiber $\psi^{-1}(\mathrm{P})$ 
for a generic point $\mathrm{P} \in \overline{\mathrm{E}}_1$.
Considering that the variables $x,y,c$ are constrained by
the curve $\overline{\mathrm{E}}_2$ one finds that such
degree is indeed four. 

An alternative way to see that
the two elliptic curves are not isomorphic is through the
comparison of their $\mathrm{J}$-invariants. It is well known
that such invariant classifies genus one curves up 
to isomorphism \cite{SILV}. This invariant can be computed
by birationally transforming a genus one curve into its
Weierstrass form, namely
\begin{equation}
\label{WEI}
\mathrm{C}=y_0^2-x_0^3-\mathbb{A}x_0-\mathbb{B},
\end{equation}
with $\mathbb{A}$ and $\mathbb{B}$ in the complex field.

Note that if we replace $x_0$ by $ \lambda^2 x_0$ and
$y_0$ by $\lambda^3 y_0$ we  still retain the main Weierstrass form of
the curve. The only amount of ambiguity
is that the coefficients $\mathbb{A}$ and
$\mathbb{B}$ are replaced by $\lambda^{-4} \mathbb{A}$
and $\lambda^{-6} \mathbb{B}$ respectively. We see that under such
scale of coordinates
there is just one invariant which is clearly the quantity
$\mathbb{A}^3/\mathbb{B}^2$. The $\mathrm{J}$-invariant
is defined as a linear fractional image of this ratio,
\begin{equation}
\mathrm{J}(\mathrm{C})= 1728 \frac{4 \mathbb{A}^3}{4\mathbb{A}^3+27\mathbb{B}^2},
\end{equation}
where the numerical prefactor is chosen for sake of compatibility
with situations in which the field characteristic is non-zero \cite{SILV}. 

The curves $\mathrm{E}_1$
and $\mathrm{E}_2$ are easily normalized to the
Weierstrass form and the 
final results 
for their $\mathrm{J}$-invariants are, 
\begin{equation}
\mathrm{J}(\mathrm{E}_1)=\frac{(\mathrm{U}^4+16\mathrm{U}^2+16)^3}{\mathrm{U}^2(\mathrm{U}^2+16)}~~\mathrm{and}~~
\mathrm{J}(\mathrm{E}_2)=-\frac{(\mathrm{U}^2+16\mathrm{U}+16)^3(\mathrm{U}^2-16\mathrm{U}+16)^3}
{\mathrm{U}^2(\mathrm{U}^2+16)^4},
\end{equation}
which are clearly different for generic values 
of $\mathrm{U}$ and consequently the
curves $\mathrm{E}_1$ and $\mathrm{E}_2$ are not isomorphic. We also note that the denominators of the $\mathrm{J}$-invariants
vanish at the the non-trivial values of the coupling $\mathrm{U}=\pm 4 \IM$ in which the curves $\mathrm{E}_1$ and $\mathrm{E}_2$
become rational.

Moreover, given two elliptic curves $\mathrm{C}_1$ and $\mathrm{C}_2$ and an integer 
$\mathrm{n}$, there is
a direct way to decide if they are $\mathrm{n}$-isogenous. We just
have to verify that the so-called modular polynomial 
$\Phi_{\mathrm{n}}\left[\mathrm{J}(\mathrm{C}_1),\mathrm{J}(\mathrm{C}_2)\right]$ is
zero. In our specific situation the expression 
of the four-level modular
polynomial is \cite{ITO},
\begin{eqnarray}
&& \Phi_4[x,y] = x^6+y^6-(x^5y^4+x^4y^5)+2976(x^5y^3+x^3y^5)-2533680(x^5y^2+x^2y^5)+561444609(x^5y+xy^5) \nonumber \\
&& -8507430000(x^5+y^5)+7440(x^4y^4)+80967606480(x^4y^3+x^3y^4)+1425220456750080(x^4y^2+x^2y^4) \nonumber \\
&&+1194227244109980000(x^4y+xy^4) + 24125474716854750000(x^4 +y^4)+2729942049541120(x^3y^3) \nonumber \\
&&-914362550706103200000(x^3y^2+x^2y^3)
+12519806366846423598750000(x^3y+xy^3) \nonumber \\
&&- 22805180351548032195000000000(x^3+y^3) 
+26402314839969410496000000(x^2y^2) \nonumber \\ 
&&+188656639464998455284287109375(x^2y+xy^2) + 158010236947953767724187500000000(x^2+y^2) \nonumber \\
&&-94266583063223403127324218750000(xy) - 364936327796757658404375000000000000(x+y) \nonumber \\
&&+ 280949374722195372109640625000000000000.
\end{eqnarray}

We have checked that the non-trivial identity
$\Phi_{4}\left[\mathrm{J}(\mathrm{E}_1),\mathrm{J}(\mathrm{E}_2)\right]=0$ is indeed satisfied 
for arbitrary values of the coupling $\mathrm{U}$. This confirms the fourfold isogeny among
the elliptic curves $\mathrm{E}_1$ and $\mathrm{E}_2$.

\section{Uniformization and local relations}
\label{SEC2}

We start showing that the uniformization of the 
curve $\overline{\mathrm{E}}_2$ can be
implemented along the lines of the eight-vertex model satisfying the
free-fermion condition \cite{BAX}. To this end we first write this elliptic
curve as the intersection of two quadric surfaces in the three-dimensional
space. Denoting by $w$ such extra coordinate, $\overline{\mathrm{E}}_2$
can be represented by the following pairs of equations,
\begin{equation}
\label{ECP4}
x^2+y^2-cw=0,~~c^2-w^2+\mathrm{U}xy=0,
\end{equation}
and after performing the rotation $c=w_1-\IM w_2$ and $w=w_1 +\IM w_2$ we
obtain,
\begin{equation}
\label{EIGH}
x^2+y^2-w_1^2-w_2^2=0,~~w_1w_2=\frac{\mathrm{U}}{4 \IM}xy.
\end{equation}

Inspecting Eq.(\ref{EIGH}) we recognize the well known
spectral curve of the symmetric eight vertex model 
with weights
$x,y,w_1$ and $w_2$ satisfying the free-fermion 
restriction. At this point
we can follow Baxter monograph \cite{BAX} and the
uniformization of the weights relevant for the 
Hubbard model are,
\begin{equation}
\frac{x(\lambda)}{c(\lambda)}=\frac{\mathrm{sn}[{\bf{K}}(k)-\lambda,k]}{1-\IM k \mathrm{sn}[\lambda,k] 
\mathrm{sn}[{\bf{K}}(k)-\lambda,k]},~~
\frac{y(\lambda)}{c(\lambda)}=\frac{\mathrm{sn}[\lambda,k]}{1-\IM k \mathrm{sn}[\lambda,k] 
\mathrm{sn}[{\bf{K}}(k)-\lambda,k]},
\end{equation}
where $\lambda$ is the spectral parameter, ${\bf{K}}(k)$ denotes the 
complete elliptic integral
of the first kind of modulus $k$ and $\mathrm{sn}[\lambda,k]$ 
represents the Jacobi elliptic
function. The dependence of the modulus on the coupling is,
\begin{equation}
k=\frac{\mathrm{U}}{4\IM}.
\end{equation}

Note that this uniformization  
when $\mathrm{U} \rightarrow 0$ recovers in a 
direct way the expected
trigonometric parameterization of the weights\footnote{ The regular point 
is at $\lambda=0$ in which
the Lax operator (\ref{LAXO}) becomes 
the four-dimensional permutator.}. This representation however involves
sums in the denominator and it is not optimal for the the study of analytical
properties. The uniformization can alternatively be given
in terms of ratios of entire functions of the spectral
parameter.
The first task is to located the positions 
and multiplicities of the
zeros and poles of the given elliptic function in the region
defined by the respective pair of primitive periods. Then we can write
the elliptic function as ratios of products of theta functions
located at such zeros and poles within a constant factor. The 
multiplicative constant can be determined by the exact knowledge
of the function at some suitable values of the spectral parameter.
Considering this procedure we find the following factorized 
representation,
\begin{eqnarray}
&& \frac{x(\lambda)}{c(\lambda)}=\IM (4 k \sqrt{\mathrm{q}})^{-1/4} 
\frac{\mathrm{H}[{\bf{K}}(k)-\lambda,k]\Theta[\lambda,k]}
{\mathrm{H}[\lambda+\IM {\bf{K}}(k^{'})/2,k]
\mathrm{H}[{\bf{K}}(k)+\IM {\bf{K}}(k^{'})/2-\lambda,k]} \\ \nonumber \\ 
&& \frac{y(\lambda)}{c(\lambda)}=\IM (4 k \sqrt{\mathrm{q}})^{-1/4} 
\frac{\Theta[{\bf{K}}(k)-\lambda,k]\mathrm{H}[\lambda,k]}
{\mathrm{H}[\lambda+\IM {\bf{K}}(k^{'})/2,k]
\mathrm{H}[{\bf{K}}(k)+\IM {\bf{K}}(k^{'})/2-\lambda,k]}
\end{eqnarray}
where the complementary modulus $k^{'}$ satisfies the 
usual relation ${k^{'}}^2+k^2=1$ and the nome 
$\mathrm{q}=\exp[-\pi {\bf{K}}(k^{'})/{\bf{K(k)}}]$. For sake of completeness the explicit expressions
of the theta functions are, 
\begin{eqnarray}
&&\mathrm{H}[\lambda,k]= 2 \mathrm{q}^{1/4} \sin \left[\frac{\pi \lambda}{2 {\bf{K}}(k)} \right] \prod_{j=1}^{\infty}
\left( 1-2 \mathrm{q}^{2j}\cos \left[ \frac{\pi \lambda}{{\bf{K}}(k)} \right] +\mathrm{q}^{4j} \right)(1-\mathrm{q}^{2j}),
\\ \nonumber \\
&& \Theta[\lambda,k]= \prod_{j=1}^{\infty}
\left( 1-2 \mathrm{q}^{(2j-1)}\cos \left[ \frac{\pi \lambda}{{\bf{K}}(k)} \right] +\mathrm{q}^{(4j-2)} \right)(1-\mathrm{q}^{2j}).
\end{eqnarray}

In order to express the Lax operator (\ref{LAXO}) solely in terms of ratios 
of entire functions we still
need the representation of the polynomial combination 
$\theta(x,y)$.  After some simplifications
it can be given as,
\begin{eqnarray}
\frac{\theta(\lambda)}{c^2(\lambda)}&=&\IM 
\frac{\Theta[{\bf{K}}(k)+\IM {\bf{K}}(k^{'})/2-\lambda,k] 
\Theta[\lambda+\IM {\bf{K}}(k^{'})/2,k]} 
{\mathrm{H}[{\bf{K}}(k)+\IM {\bf{K}}(k^{'})/2-\lambda,k] 
\mathrm{H}[\lambda+\IM {\bf{K}}(k^{'})/2,k]} \nonumber \\ 
&=&\IM \frac{\Theta[{\bf{K}}(k)+\IM {\bf{K}}(k^{'})/2-\lambda,k] 
\mathrm{H}[\lambda-\IM {\bf{K}}(k^{'})/2,k]} 
{\mathrm{H}[{\bf{K}}(k)+\IM {\bf{K}}(k^{'})/2-\lambda,k] 
\Theta[\lambda-\IM {\bf{K}}(k^{'})/2,k]} 
\end{eqnarray}

We have now gathered the basic ingredients to discuss local 
properties satisfied by the Lax operator. One of them is related with 
the invariance of the respective partition function by
$\pi/2$ rotation of the lattice. Inspecting the structure
of the operator (\ref{LAXO}) we conclude that this symmetry is
directly related with the variables exchange 
$x \leftrightarrow y$ which
preserves the form of the spectral curve (\ref{CURVEP2}). 
Considering the above uniformization we see that this exchange is
accomplished by shifting the spectral parameter by the
elliptic integral ${\bf{K}}(k)$ value. Denoting by 
$\mathrm{L}_{12}(\lambda)$
the ratio $\mathrm{L}_{12}(x,y,c)/c^2$ we
found the following crossing relation,
\begin{equation}
\label{cross}
\mathrm{L}_{12}(\lambda)=\mathrm{M}_1 \mathrm{L}_{12}({\bf{K}}(k)-\lambda)^{t_2} \mathrm{M}_1^{-1},
\end{equation}
where $t_2$ denotes transposition on the second space and the
charge conjugation matrix $\mathrm{M}$ is,
\begin{equation}
\mathrm{M}=\left(
\begin{array}{cccc}
0& 0& 0& 1 \\
0& 0& 1& 0 \\
0& 1& 0& 0 \\
1& 0& 0& 0 \\
\end{array}
\right).
\end{equation}

The crossing property has the immediate consequence of providing us a global 
symmetry constrain for the free energy of the classical vertex model at finite volume.
Let $\mathrm{Z}_{\mathrm{N}}(\lambda)$ be the partition function
of the vertex model with weights $\mathrm{L}_{12}(\lambda)$ on the
square lattice of size $\mathrm{N}$. Then it follows from Eq.(\ref{cross}) that,
\begin{equation}
\mathrm{Z}_{\mathrm{N}}(\lambda)=
\mathrm{Z}_{\mathrm{N}}\left({\bf{K}}(k)-\lambda\right)
\end{equation}
 
The next local property is the so-called unitarity relation which for
relativistic scattering theory connects
Lax operators with spectral parameters $\lambda$ and $-\lambda$. Here
we have attempted similar relation by studying the local properties 
of the Lax operator
around the regular point $\lambda=0$. The result of this analysis is the following expression,
\begin{equation}
\mathrm{L}_{12}(\lambda)\mathrm{L}_{12}(-\lambda)= \left[\frac{x(\lambda)}{c(\lambda)}\right]^2
\left[\frac{x(-\lambda)}{c(-\lambda)}\right]^2 \mathrm{I}_1 \otimes \mathrm{I}_2
\end{equation}

Note that the above relation is almost what we usually have 
for relativistic systems. The only
difference is that the Lax operator evaluated at $-\lambda$ is not permuted 
on its spaces. Here the Lax operator
is not parity reversal invariant for generic values of $\mathrm{U}$ and 
as a consequence of that $\mathrm{L}_{12}(\lambda) \neq
\mathrm{L}_{21}(\lambda)$. From previous experience with other 
solvable models
it is conceivable that combination of crossing and unitarity could lead us
to functional relations for the transfer matrix eigenvalues in the limit
of infinite system \cite{STRO,BAX1}. This method provides an alternative way to 
derive relevant physical properties such as the free-energy and the
dispersion relation of the low-lying excitations \cite{KLU}. We hope
that our theta function uniformization of the weights will shed some light 
on the appropriate
analyticity assumptions that still has to be made 
for the applicability of such approach.

\section{$\mathrm{R}$-matrix geometric properties }
\label{SEC3}
We start this section by presenting the explicit 
expression of Shastry's $\mathrm{R}$-matrix 
in terms of the suitable ring variables describing the 
Lax operator. This matrix can be
written as,
\begin{equation}
\mathrm{R}(\lambda_1,\lambda_2)=\left(
\begin{array}{cccccccccccccccc}
\bf{a}& 0& 0& 0& 0& 0& 0& 0& 0& 0& 0& 0& 0& 0& 0& 0 \\
0& \bf{b}& 0& 0& \bf{c}& 0& 0& 0& 0& 0& 0& 0& 0& 0& 0& 0 \\
0& 0& \bf{b}& 0& 0& 0& 0& 0& \bf{c}& 0& 0& 0& 0& 0& 0& 0 \\
0& 0& 0& \bf{h-a}& 0& 0& \bf{d}& 0& 0& \bf{d}& 0& 0& \bf{h}& 0& 
0& 0 \\
0& \bf{c}& 0& 0& \overline{\bf{b}} & 0& 0& 0& 0& 
      0& 0& 0& 0& 0& 0& 0 \\
0& 0& 0& 0& 0& \bf{g} & 0& 
      0& 0& 0& 0& 0& 0& 0& 0& 0 \\
0& 0& 0& \bf{d} & 0& 0& 
      \bf{q}-\bf{g}& 0& 0& \bf{q} & 0& 0& \bf{d} & 0& 0& 0 \\
     0& 0& 0& 0& 0& 0& 0& \bf{b} & 0& 0& 0& 0& 0& 
      \bf{c} & 0& 0 \\
0& 0& \bf{c} & 0& 0& 0& 0& 0& 
      \overline{\bf{b}} & 0& 0& 0& 0& 0& 0& 0 \\
     0& 0& 0& \bf{d} & 0& 0& \bf{q} & 0& 0& \bf{q}-\bf{g} & 0& 
      0& \bf{d} & 0& 0& 0 \\
0& 0& 0& 0& 0& 0& 0& 0& 0& 0& 
      \bf{g} & 0& 0& 0& 0& 0 \\ 
     0& 0& 0& 0& 0& 0& 0& 0& 0& 0& 0& \overline{\bf{b}} & 0& 0& 
      \bf{c} & 0 \\
0& 0& 0& \bf{h} & 0& 0& \bf{d} & 0& 0& \bf{d} & 0& 
      0& \bf{h}-\bf{a} & 0& 0& 0 \\
0& 0& 0& 0& 0& 0& 0& \bf{c} & 0& 0& 0& 0& 0& 
      \bf{b} & 0& 0 \\
0& 0& 0& 0& 0& 0& 0& 0& 0& 0& 0& \bf{c} & 0& 0& 
      \bf{b} & 0 \\
0& 0& 0& 0& 0& 0& 0& 0& 0& 0& 0& 0& 0& 0& 0& \bf{a} \\
\end{array}
\right),
\end{equation}

The expressions for the $\mathrm{R}$-matrix elements 
are obtained after performing some
simplifications  
on the original weights determined previously by Shastry \cite{SHA1,SHA2}. Considering
the entry $\bf{c}$ as an overall normalization we found,
\begin{eqnarray}
\label{weigR}
&& \frac{\bf{a}}{\bf{c}}=\frac{{\bf{y}}_1{\bf{y}}_2}{\theta({\bf{x}}_1,{\bf{y}}_1)}
+\frac{{\bf{x}}_1{\bf{x}}_2}{\theta({\bf{x}}_2,{\bf{y}}_2)},~~
\frac{\bf{b}}{\bf{c}}=-\frac{{\bf{x}}_1{\bf{y}}_2}{\theta({\bf{x}}_1,{\bf{y}}_1)}
+\frac{{\bf{y}}_1{\bf{x}}_2}{\theta({\bf{x}}_2,{\bf{y}}_2)},~~
\frac{\overline{\bf{b}}}{\bf{c}}=\frac{{\bf{y}}_1{\bf{x}}_2}{\theta({\bf{x}}_1,{\bf{y}}_1)}
-\frac{{\bf{x}}_1{\bf{y}}_2}{\theta({\bf{x}}_2,{\bf{y}}_2)},\nonumber \\ \nonumber \\ 
&&\frac{\bf{d}}{\bf{c}}=\frac{{\bf{x}}_1{\bf{y}}_1-{\bf{x}}_2{\bf{y}}_2}
{{\bf{x}}_1^2{\bf{x}}_2^2-{\bf{y}}_1^2{\bf{y}}_2^2},~~
\frac{\bf{h}}{\bf{c}}=\frac{{\bf{x}}_1{\bf{x}}_2\theta({\bf{x}}_1,{\bf{y}}_1)-{\bf{y}}_1{\bf{y}}_2\theta({\bf{x}}_2,{\bf{y}}_2)}{{\bf{x}}_1^2{\bf{x}}_2^2-{\bf{y}}_1^2{\bf{y}}_2^2}, \nonumber \\ \nonumber \\
&&\frac{\bf{q}}{\bf{c}}=\frac{{\bf{x}}_1{\bf{x}}_2\theta({\bf{x}}_2,{\bf{y}}_2)-{\bf{y}}_1{\bf{y}}_2\theta({\bf{x}}_1,{\bf{y}}_1)}{{\bf{x}}_1^2{\bf{x}}_2^2-{\bf{y}}_1^2{\bf{y}}_2^2},~~
\frac{\bf{g}}{\bf{c}}=\frac{{\bf{x}}_1{\bf{x}}_2}{\theta({\bf{x}}_1,{\bf{y}}_1)}
+\frac{{\bf{y}}_1{\bf{y}}_2}{\theta({\bf{x}}_2,{\bf{y}}_2)},
\end{eqnarray}
where the bold variables 
${\bf{x}}_j$ and 
${\bf{y}}_j$ are given as ratios of the 
spectral curve coordinates,
\begin{equation}
{\bf{x}}_j=\frac{x(\lambda_j)}{c(\lambda_j)},~~
{\bf{y}}_j=\frac{y(\lambda_j)}{c(\lambda_j)},~~\mathrm{for}~~j=1,2.
\end{equation}

In order to understand the geometric properties 
associated to the $\mathrm{R}$-matrix
we first need to find the implicit representation of 
the image of the rational map,
\EQ
\label{mapR}
\renewcommand{\arraystretch}{1.5}
\begin{array}{ccc}
\overline{\mathrm{E}}_2 \times \overline{\mathrm{E}}_2 \subset \mathbb{CP}^2 \times \mathbb{CP}^2 &~~~ \overset{\phi}{\longrightarrow}~~~ 
& \mathrm{V} \subset \mathbb{CP}^7 \\
\left(x(\lambda_1):y(\lambda_1):c(\lambda_1)\right) \times
\left(x(\lambda_2):y(\lambda_2):c(\lambda_2)\right)
& \longmapsto &~(\bf{a}:\bf{b}:\overline{\bf{b}}:\bf{c}:\bf{d}:\bf{g}:\bf{h}:\bf{q}),
\end{array}
\EN
where $\mathrm{V}$ is the algebraic variety associated to the $\mathrm{R}$-matrix.

The solution of the above problem will 
lead us to polynomials 
on the $\mathrm{R}$-matrix entries 
$\bf{a},\bf{b},\overline{\bf{b}},\bf{c},\bf{d},\bf{g},\bf{h}$ and $\bf{q}$
which are the defining equations of $\mathrm{V}$. This task is performed by
eliminating the variables ${\bf{x}}_j$ and ${\bf{y}}_j$ 
from Eqs.(\ref{weigR}) considering also that they are
constrained by the spectral curve (\ref{CURVE2}). The technical details concerning
this computation are summarized in Appendix A and in what follows we only 
present the main results. It turns out that the variety $\mathrm{V}$ is formally
described as the intersection of five quadrics,
\begin{equation}
\mathrm{V}=\{
({\bf{a}}:{\bf{b}}:\overline{\bf{b}}:{\bf{c}}:{\bf{d}}:{\bf{g}}:{\bf{h}}:{\bf{q}}
) \in \mathbb{CP}^7 |\mathrm{Q}_1=\mathrm{Q}_2=\mathrm{Q}_3=\mathrm{Q}_4=\mathrm{Q}_5=0 \},
\end{equation}

The 
expressions of the degree two homogeneous polynomials $\mathrm{Q}_j$ are,
\begin{eqnarray}
&&\mathrm{Q}_1 = -{\bf{c}}^2 + {\bf{a}}{\bf{g}} + {\bf{b}}\overline{\bf{b}},~~
\mathrm{Q}_2 = -{\bf{d}}^2 + {\bf{a}}{\bf{g}} - {\bf{g}}{\bf{h}} 
- {\bf{a}}{\bf{q}} + {\bf{h}}{\bf{q}} + {\bf{b}}\overline{\bf{b}}~~,
\mathrm{Q}_3 = -{\bf{c}}^2 - {\bf{d}}^2 + {\bf{h}}{\bf{q}}, \nonumber \\
&&\mathrm{Q}_4 =-{\bf{a}}^2 - {\bf{b}}^2 - {\bf{g}}^2 + {\bf{a}}{\bf{h}} + {\bf{g}}{\bf{q}} - \overline{\bf{b}}^2,~~
\mathrm{Q}_5 = \mathrm{U}{\bf{c}}{\bf{d}} - {\bf{h}}^2 + {\bf{q}}^2
\end{eqnarray}
where we recall
that the above first three quadrics have been pointed before as identities among
the $\mathrm{R}$-matrix weights in \cite{WA}. However, to the best of 
our knowledge the last two are new 
in the literature specially
$\mathrm{Q}_5$ since it contains the Hubbard coupling $\mathrm{U}$. 

We have used the computer algebra system Singular \cite{SIN} to obtain 
some basic information
on the geometric properties of the variety $\mathrm{V}$. This algebraic set turns out 
to be an irreducible complete intersection and therefore we are dealing with a complex
two-dimensional variety. This distinguishes the Hubbard and the eight-vertex
models even though both have Lax operator based on elliptic curves. 
In fact, for the eight-vertex
model the variety $\mathrm{V}$ is one-dimensional and the $\mathrm{R}$-matrix
lies on the same curve of the Lax operator \cite{BAX} and the map (\ref{mapR}) reflects the
standard group law of elliptic curves. By way of contrast, the Hubbard model sits on the lower
bound of the fiber dimension theorem\footnote{This theorem states that if 
$\phi: \mathrm{X} \rightarrow \mathrm{Y}$ is a surjective morphism among 
irreducible varieties then
$\mathrm{dim(\phi^{-1})} \geq \mathrm{dim(X)} -\mathrm{dim(Y)}$, see for example \cite{SHAFA}.}
in which $\phi^{-1}$ is a zero dimensional variety.

Further progress is made by noticing that the quadrics 
$\mathrm{Q}_3$ and $\mathrm{Q}_5$ define
a nonsingular elliptic curve in $\mathbb{CP}^3[{\bf{c}},{\bf{d}},{\bf{h}},{\bf{q}}]$ which
is isomorphic to $\overline{\mathrm{E}}_2$ formulated as in Eq.(\ref{ECP4}).
This means that $\mathrm{V}$ is a surface contained in the cone with base  
$\mathbb{CP}^3[{\bf{a}},{\bf{b}},\overline{{\bf{b}}},{\bf{g}}]$ 
over $\overline{\mathrm{E}}_2$ making it possible to 
established the following surjective map,
\EQ
\label{mapF}
\renewcommand{\arraystretch}{1.5}
\begin{array}{ccc}
\mathrm{V} \subset \mathbb{CP}^7 &~~~ \overset{\pi} {\longrightarrow}~~~ 
& \overline{\mathrm{E}}_2 \subset \mathbb{CP}^3 \\
(\bf{a}:\bf{b}:\overline{\bf{b}}:\bf{c}:\bf{d}:\bf{g}:\bf{h}:\bf{q})
& \longmapsto &~({\bf{c}}:{\bf{d}}:{\bf{h}}:{\bf{q}}),
\end{array}
\EN

The next natural step is to investigate the  properties
of the fiber of $\pi$ since this feature lies at the heart of the
geometry of algebraic surfaces \cite{BU}. 
This study is somehow cumbersome and the main technical 
points of the
computations have been
summarized in Appendix B. The central result of this analysis is that
the general fiber $\pi^{-1}$ turns out to be a smooth 
curve of genus one meaning that $\mathrm{V}$ is 
an elliptic surface. From the classification 
theory of algebraic
surfaces \cite{BU} we know that an elliptic surface fibred
over a genus one curve can be 
either an Abelian surface, a bielliptic
surface or a proper elliptic surface with Kodaira dimension
one. In order to decide on the actual class of $\mathrm{V}$ a 
central ingredient is the description of the generic fiber
in terms of its Weierstrass model. Now the respective 
pair of coefficients $\mathbb{A}$ and $\mathbb{B}$ are
interpreted as local functions on the curve
$\overline{\mathrm{E}}_2$ and from this data we shall be able
to infer on the class of the surface. We have found that
the equation for such Weierstrass fibration has a remarkable 
simple structure, namely
\begin{equation}
\label{WEIF}
y_0^2-x_0^3
+\frac{c_0^4d_0^4(\mathrm{U}^4+246\mathrm{U}^2+4096)}{48}x_0
+\frac{c_0^6d_0^6(32+\mathrm{U}^2)(\mathrm{U}^4-512\mathrm{U}^2-8192)}{864}=0,
\end{equation}
where $c_0$ and $d_0$ are coordinates of the affine point 
$[c_0,d_0,c_0^2+d_0^2,1] \subset \overline{\mathrm{E}}_2$. For the
explicit birational map dependence of $x_0$ and $y_0$ with the
surface variables see Appendix B.

We now can just change coordinates replacing $x_0$ by $x_0c_0^2d_0^2$
and $y_0$ by $y_0c_0^3d_0^3$ and dividing through $c_0^6d_0^6$ we end up
with coefficients not depending on $\overline{\mathrm{E}}_2$.
This
means that locally the Weierstrass fibration can always be definable  
with constants $\mathbb{A}$ and $\mathbb{B}$ and therefore we conclude
that $\mathrm{V}$ is an Abelian surface. More precisely, this surface
is birational to the product of two elliptic curves, namely
\begin{equation}
\mathrm{V} \cong \overline{\mathrm{E}}_2 \times \overline{\mathrm{E}}_3,
\end{equation}
where $\overline{\mathrm{E}}_3$ is defined by the homogeneous
polynomial,
\begin{equation}
\overline{\mathrm{E}}_3 \equiv z_0y_0^2-x_0^3
+\frac{(\mathrm{U}^4+246\mathrm{U}^2+4096)}{48}x_0z_0^2
+\frac{(32+\mathrm{U}^2)(\mathrm{U}^4-512\mathrm{U}^2-8192)}{864}z_0^3=0.
\end{equation}

At this point we observe that the elliptic curves
$\overline{\mathrm{E}}_2$  and $ \overline{\mathrm{E}}_3$ 
are not isomorphic but only have a degree four isogeny. In fact,
the $\mathrm{J}$-invariant of 
$\overline{\mathrm{E}}_3$ is,
\begin{equation}
\label{JIN3}
\mathrm{J}(\mathrm{E}_3)=\frac{(\mathrm{U}^4+256\mathrm{U}^2+4096)^3}{\mathrm{U}^8(\mathrm{U}^2+16)},
\end{equation}
such that it satisfies the modular
$\Phi_{4}\left[\mathrm{J}(\mathrm{E}_2),\mathrm{J}(\mathrm{E}_3)\right]=0$ identity.

The above analysis explain why the $\mathrm{R}$-matrix associated to the Hubbard can not be
written solely in terms of the difference of two spectral parameters. Besides having weights lying
on a non-trivial surface only part of its geometry retains isomorphism with the one of the
Lax operator. 

\section{Conclusions}
\label{SEC4}

The basic ingredients in the theory of solvable two-dimensional vertex model of 
statistical mechanics are the Lax operator and the $\mathrm{R}$-matrix which are
constrained by the Yang-Baxter equation (\ref{YBAX}). The Lax operator
is expected to leave on some algebraic variety $\mathrm{X}$ while the $\mathrm{R}$-matrix
may generically be sitting on a yet another manifold $\mathrm{Y}$. They may coincide in some
special situations such as when both the Lax operator and the $\mathrm{R}$-matrix
are equidimensional and
invariant by parity-time reversal symmetry. In fact, taking the transposition on
the three spaces of Eq.(\ref{YBAX}) we obtain
\begin{equation}
\mathrm{L}_{23}(\omega_2)^{t_2t_3} \mathrm{L}_{13}(\omega_1)^{t_1t_3} \mathrm{R}_{12}(\omega_1,\omega_2)^{t_1t_2}
=\mathrm{R}_{12}(\omega_1,\omega_2)^{t_1t_2} \mathrm{L}_{13}(\omega_1)^{t_1t_3} \mathrm{L}_{23}(\omega_2)^{t_2t_3},
\end{equation}
and after assuming $\mathrm{PT}$ symmetry for both operators we have,
\begin{equation}
\label{YBAX1}
\mathrm{L}_{32}(\omega_2) \mathrm{L}_{31}(\omega_1) \mathrm{R}_{21}(\omega_1,\omega_2)
=\mathrm{R}_{21}(\omega_1,\omega_2)\mathrm{L}_{31}(\omega_1) \mathrm{L}_{32}(\omega_2).
\end{equation}

Now by applying the permutation on the first and third spaces on both sides of Eq.(\ref{YBAX1})
we finally find,
\begin{equation}
\mathrm{L}_{12}(\omega_2) \mathrm{L}_{13}(\omega_1) \mathrm{R}_{12}(\omega_1,\omega_2)
=\mathrm{R}_{23}(\omega_1,\omega_2)\mathrm{L}_{13}(\omega_1) \mathrm{L}_{12}(\omega_2),
\end{equation}
and direct comparison with the original relation (\ref{YBAX}) tells us
that we have just exchanged the second Lax operator with the 
$\mathrm{R}$-matrix. This means that both $\mathrm{X}$ and $\mathrm{Y}$ should
be defined by the same polynomial relations.

In general situations the Yang-Baxter offers us a rational map since  
the $\mathrm{R}$-matrix elements can be linearly eliminated from a 
subset of independent
functional relations.
Formally this map can be represented as,
\EQ
\renewcommand{\arraystretch}{1.5}
\begin{array}{ccc}
\mathrm{X} \times \mathrm{X} \subset \mathbb{CP}^{n+1} \times \mathbb{CP}^{n+1} &~~~ \overset{\phi}{\longrightarrow}~~~ 
& \mathrm{Y} \subset \mathbb{CP}^{m} \\
(x_0:\cdots:x_{n+1}) \times
(y_0:\cdots:y_{n+1}) 
& \longmapsto &~(\phi_0(x_0,\cdots,y_{n+1}):\cdots:\phi_m(x_0,\cdots,y_{n+1}),
\end{array}
\EN
where $n=\mathrm{dim(X)}$, $m$ counts the number of 
linearly independent $\mathrm{R}$-matrix weights and 
$\phi_j(x_0,\cdots,y_{n+1})$ are map polynomials.

The study of the geometric properties of $\mathrm{Y}$ requires the implicit 
representation of the image of the map $\phi$. This is basically an elimination
problem and in principle can be solved by methods based on Gr\"obner basis
computations. In practice however it is known that this is not a simple task 
depending much on the number and complexity of the 
polynomials $\phi_j(x_0,\cdots,y_{n+1})$ 
as well as on the defining equations of $\mathrm{X}$.

In this paper we have addressed these problems 
for the classical vertex model associated
to the Hubbard Hamiltonian devised by Shastry \cite{SHA,SHA1,SHA2}. 
We find that the variety $\mathrm{X}$ 
is a genus one
curve and provided its uniformization in terms of 
factorized ratios of theta functions. This
pave the way to discuss local relations for the Lax operator much like
in the case of relativistic systems. On the other hand the geometric properties of
$\mathrm{Y}$ is that of an Abelian surface birational to the
product of two non isomorphic elliptic curves. This may explain why the Bethe
ansatz equations of the Hubbard model is somehow unconventional as compared with
other Lattice models based on elliptic curves such as the 
eight-vertex and hard-hexagon models \cite{BAX,PER}. In the algebraic Bethe 
ansatz much of the 
input comes from the $\mathrm{R}$-matrix elements which here sits in a different
algebraic variety of the respective Lax operator. It seems interesting
to look
for alternative solutions for the transfer matrix spectrum more
based on the properties of the Lax operator such as to establish
finite system exact inversion identities. 
In this context an earlier attempt by Shastry himself \cite{SHA2} and the recent 
formulation of fusion for integrable models
with $\mathrm{R}$-matrix without the difference form \cite{BEMA} 
could be relevant guidelines. We hope that the uniformization
given here will be useful for setting up this approach and 
the needed analyticity
assumptions.

\section*{Acknowledgments}
This work has been partially supported by the Brazilian Research Agencies 
FAPESP and CNPq. I am grateful to Niklas Beisert for 
fruitful discussions and the hospitality at the Institute for Theoretical Physics,
Zurich, where part of this work has been written.

\addcontentsline{toc}{section}{Appendix A}
\section*{\bf Appendix A: Elimination Procedure}
\setcounter{equation}{0}
\renewcommand{\theequation}{A.\arabic{equation}}
We start by defining the ideal $\mathrm{I} \subset \mathbb{C}[ 
{\bf{x}}_1,{\bf{y}}_1,{\bf{x}}_2,{\bf{y}}_2 
,{\bf{a}},{\bf{b}},\overline{\bf{b}},{\bf{c}},
{\bf{d}},{\bf{g}},{\bf{h}},{\bf{q}}]$ associated 
the map (\ref{mapR}) by clearing the
denominators of Eqs.(\ref{weigR}). This can be done by choosing 
appropriately the weight 
${\bf{c}}$ and as result we obtain,
\begin{eqnarray}
&& \mathrm{I}=< \mathrm{E}_2({\bf{x}}_1,{\bf{y}}_1),\mathrm{E}_2({\bf{x}}_2,{\bf{y}}_2),
{\bf{a}}-p_1({\bf{x}}_1,{\bf{y}}_1,{\bf{x}}_2,{\bf{y}}_2),
{\bf{b}}-p_2({\bf{x}}_1,{\bf{y}}_1,{\bf{x}}_2,{\bf{y}}_2),
\overline{\bf{b}}-p_3({\bf{x}}_1,{\bf{y}}_1,{\bf{x}}_2,{\bf{y}}_2), \nonumber \\
&& {\bf{c}}-p_4({\bf{x}}_1,{\bf{y}}_1,{\bf{x}}_2,{\bf{y}}_2),
{\bf{d}}-p_5({\bf{x}}_1,{\bf{y}}_1,{\bf{x}}_2,{\bf{y}}_2),
{\bf{g}}-p_6({\bf{x}}_1,{\bf{y}}_1,{\bf{x}}_2,{\bf{y}}_2), 
{\bf{h}}-p_7({\bf{x}}_1,{\bf{y}}_1,{\bf{x}}_2,{\bf{y}}_2), \nonumber \\
&& {\bf{q}}-p_8({\bf{x}}_1,{\bf{y}}_1,{\bf{x}}_2,{\bf{y}}_2)>, 
\end{eqnarray}
where the symbol $\mathrm{E}_2(x_j,y_j)$ denotes the curve 
(\ref{CURVE2}) on the variables $x_j$ and $y_j$ and
the expressions for the polynomials 
$p_j({\bf{x}}_1,{\bf{y}}_1,{\bf{x}}_2,{\bf{y}}_2)$ are,
\begin{eqnarray}
&&p_1({\bf{x}}_1,{\bf{y}}_1,{\bf{x}}_2,{\bf{y}}_2)=\left[{\bf{y}}_1 {\bf{y}}_2 \theta({\bf{x}}_2,{\bf{y}}_2)+{\bf{x}}_1 {\bf{x}}_2 \theta({\bf{x}}_1,{\bf{y}}_1)\right]
\left[{\bf{x}}_1^2{\bf{x}}_2^2-{\bf{y}}_1^2{\bf{y}}_2^2\right], \nonumber \\
&&p_2({\bf{x}}_1,{\bf{y}}_1,{\bf{x}}_2,{\bf{y}}_2)=\left[{\bf{y}}_1{\bf{x}}_2\theta({\bf{x}}_1,{\bf{y}}_1)-{\bf{x}}_1{\bf{y}}_2\theta({\bf{x}}_2,{\bf{y}}_2)\right] \left[{\bf{x}}_1^2{\bf{x}}_2^2-{\bf{y}}_1^2{\bf{y}}_2^2\right], \nonumber \\
&&p_3({\bf{x}}_1,{\bf{y}}_1,{\bf{x}}_2,{\bf{y}}_2)=\left[{\bf{y}}_1{\bf{x}}_2\theta({\bf{x}}_2,{\bf{y}}_2)-{\bf{x}}_1{\bf{y}}_2\theta({\bf{x}}_1,{\bf{y}}_1)\right]\left[{\bf{x}}_1^2{\bf{x}}_2^2-{\bf{y}}_1^2{\bf{y}}_2^2\right], \nonumber \\
&&p_4({\bf{x}}_1,{\bf{y}}_1,{\bf{x}}_2,{\bf{y}}_2)=\theta({\bf{x}}_1,{\bf{y}}_1)\theta({\bf{x}}_2,{\bf{y}}_2)\left[{\bf{x}}_1^2{\bf{x}}_2^2-{\bf{y}}_1^2{\bf{y}}_2^2\right], \nonumber \\
&& p_5({\bf{x}}_1,{\bf{y}}_1,{\bf{x}}_2,{\bf{y}}_2)=\left[{\bf{x}}_1{\bf{y}}_1-{\bf{x}}_2{\bf{y}}_2\right]\theta({\bf{x}}_1,{\bf{y}}_1)\theta({\bf{x}}_2,{\bf{y}}_2), \nonumber \\
&&p_6({\bf{x}}_1,{\bf{y}}_1,{\bf{x}}_2,{\bf{y}}_2)=\left[{\bf{x}}_1{\bf{x}}_2\theta({\bf{x}}_2,{\bf{y}}_2)+{\bf{y}}_1{\bf{y}}_2\theta({\bf{x}}_1,{\bf{y}}_1)\right]\left[{\bf{x}}_1^2{\bf{x}}_2^2-{\bf{y}}_1^2{\bf{y}}_2^2\right], \nonumber \\
&& p_7({\bf{x}}_1,{\bf{y}}_1,{\bf{x}}_2,{\bf{y}}_2)=\left[{\bf{x}}_1{\bf{x}}_2\theta({\bf{x}}_1,{\bf{y}}_1)-{\bf{y}}_1{\bf{y}}_2\theta({\bf{x}}_2,{\bf{y}}_2)\right]\theta({\bf{x}}_1,{\bf{y}}_1)\theta({\bf{x}}_2,{\bf{y}}_2), \nonumber \\
&& p_8({\bf{x}}_1,{\bf{y}}_1,{\bf{x}}_2,{\bf{y}}_2)=\left[{\bf{x}}_1{\bf{x}}_2\theta({\bf{x}}_2,{\bf{y}}_2)-{\bf{y}}_1{\bf{y}}_2\theta({\bf{x}}_1,{\bf{y}}_1)\right]\theta({\bf{x}}_1,{\bf{y}}_1)\theta({\bf{x}}_2,{\bf{y}}_2).
\end{eqnarray}

The elimination of the variables 
${\bf{x}}_1,{\bf{y}}_1,{\bf{x}}_2,{\bf{y}}_2$ of the above polynomials 
is equivalent to find
the ideal $\mathrm{I}_1 \subset \mathbb{C}
[{\bf{a}},{\bf{b}},\overline{\bf{b}},{\bf{c}},
{\bf{d}},{\bf{g}},{\bf{h}},{\bf{q}}]$ defined by,
\begin{equation}
\mathrm{I}_1 = \mathrm{I} \cap \mathbb{C}
[{\bf{a}},{\bf{b}},\overline{\bf{b}},{\bf{c}},
{\bf{d}},{\bf{g}},{\bf{h}},{\bf{q}}].
\end{equation}

One way of finding $\mathrm{I}_1$ is first 
to compute an alternative
basis of $\mathrm{I}$ called Gr\"obner basis. The elimination 
theorem asserts that if $\mathrm{G}$ is the Gr\"obner basis of 
$\mathrm{I}$ then  
$\mathrm{G} \cap \mathbb{C}
[{\bf{a}},{\bf{b}},\overline{\bf{b}},{\bf{c}},
{\bf{d}},{\bf{g}},{\bf{h}},{\bf{q}}]$ 
is a Gr\"obner basis of $\mathrm{I}_1$. For more details about
this theorem and its properties we refer to the Book \cite{COX}.
Fortunately all that can
be computed using intrinsics developed in some computer algebra
systems such as Singular \cite{SIN}. Direct computations are however
involved and we find more convenient to eliminate each pair of variables
${\bf{x}}_j,{\bf{y}}_j$ at a time. It turns out that the
elimination of the variables 
${\bf{x}}_1$ and ${\bf{y}}_1$ leads to an intermediate ideal 
$\mathrm{I}_2 \subset \mathbb{C}[
{\bf{x}}_2,{\bf{y}}_2 
,{\bf{a}},{\bf{b}},\overline{\bf{b}},{\bf{c}},
{\bf{d}},{\bf{g}},{\bf{h}},{\bf{q}}]$  
whose generating set of polynomials are given by,
\begin{eqnarray}
&& \mathrm{I}_2^{(1)}={\bf{a}}{\bf{g}}-{\bf{c}}^2+{\bf{b}}\overline{{\bf{b}}}, \nonumber \\
&&\mathrm{I}_2^{(2)}=({\bf{b}}^2+\overline{{\bf{b}}}^2+{\bf{a}}^2-{\bf{a}}{\bf{h}})({\bf{h}}-{\bf{a}})^3 +{\bf{a}}({\bf{d}}^2-{\bf{b}}\overline{{\bf{b}}})^2-2{\bf{b}}\overline{{\bf{b}}}({\bf{h}}-{\bf{a}})({\bf{d}}^2-{\bf{b}}\overline{{\bf{b}}}),
\nonumber \\
&& \mathrm{I}_2^{(3)}\equiv \mathrm{E}_2({\bf{x}}_2,{\bf{y}_2})= ({\bf{x}}_2^2+{\bf{y}}_2^2)^2-\mathrm{U}{\bf{x}_2}{\bf{y}_2}-1, \nonumber \\
&& \mathrm{I}_2^{(4)}={\bf{b}}^2+{\bf{a}}^2-{\bf{a}}{\bf{h}}+\omega_1({\bf{x}}_2,{\bf{y}}_2){\bf{c}}{\bf{d}}, \nonumber \\
&& \mathrm{I}_2^{(5)}={\bf{b}}{\bf{c}}+\omega_1({\bf{x}}_2,{\bf{y}}_2)\overline{{\bf{b}}}{\bf{d}}-\omega_2({\bf{x}}_2,{\bf{y}}_2){\bf{a}}{\bf{d}}, \nonumber \\
&& \mathrm{I}_2^{(6)}=\omega_2({\bf{x}}_2,{\bf{y}}_2){\bf{b}}{\bf{d}}+\omega_1({\bf{x}}_2,{\bf{y}}_2)\omega_2({\bf{x}}_2,{\bf{y}}_2)\overline{{\bf{b}}}{\bf{c}}-\left[1+\omega_1({\bf{x}}_2,{\bf{y}}_2)^2\right]({\bf{h}}-{\bf{a}}){\bf{c}}, \nonumber \\
&& \mathrm{I}_2^{(7)}=\omega_2({\bf{x}}_2,{\bf{y}}_2){\bf{a}}{\bf{d}}-\omega_1({\bf{x}}_2,{\bf{y}}_2)\omega_2({\bf{x}}_2,{\bf{y}}_2)({\bf{q}}-{\bf{g}}){\bf{c}}-\left[1+\omega_1({\bf{x}}_2,{\bf{y}}_2)^2\right]{\bf{b}}{\bf{c}},
\end{eqnarray}
where we recognize that the first component $\mathrm{I}_2^{(1)}$ is exactly the quadratic $\mathrm{Q}_1$. The
functions depending on the variables ${\bf{x}}_2$ and ${\bf{y}}_2$ are,
\begin{equation}
\omega_1({\bf{x}}_2,{\bf{y}}_2)=
\frac{\mathrm{U}{\bf{x}}_2^2{\bf{y}}_2^2}{\mathrm{U}{\bf{x}}_2{\bf{y}}_2+1},~~
\omega_2({\bf{x}}_2,{\bf{y}}_2)={\bf{x}}_2^2+
\frac{{\bf{y}}_2^2}{\mathrm{U}{\bf{x}}_2{\bf{y}}_2+1}
\end{equation}

We now proceed by eliminating the fraction field elements 
$\omega_1({\bf{x}}_2,{\bf{y}}_2)$ and 
$\omega_2({\bf{x}}_2,{\bf{y}}_2)$ out of the generators 
$\mathrm{I}_2^{(3)}, \cdots, \mathrm{I}_2^{(7)}$. The
compatibility between $\mathrm{I}_2^{(6)}$ and $\mathrm{I}_2^{(7)}$ leads us directly
to the quadratic $\mathrm{Q}_2$ as well as to the following polynomial,
\begin{equation}
\mathrm{I}_2^{(6)}=({\bf{c}}^2-{\bf{b}}\overline{{\bf{b}}})({\bf{d}}^2-{\bf{b}}\overline{{\bf{b}}})+{\bf{a}}({\bf{h}}-{\bf{a}})\left({\bf{a}}({\bf{h}}-{\bf{a}})-{\bf{b}}^2-\overline{{\bf{b}}}^2\right)
\end{equation}

It turns out that the  above generator can be further simplified with the help of the quadrics 
$\mathrm{Q}_1$ and $\mathrm{Q}_2$, namely
\begin{eqnarray}
\mathrm{I}_6^{(2)}&=&{\bf{a}}{\bf{g}}({\bf{h}}-{\bf{a}})({\bf{q}}-{\bf{g}})+{\bf{a}}({\bf{h}}-{\bf{a}})\left({\bf{a}}({\bf{h}}-{\bf{a}})-{\bf{b}}^2-\overline{{\bf{b}}}^2\right) \nonumber \\
&=& {\bf{a}}({\bf{h}}-{\bf{a}})\left[{\bf{g}}({\bf{q}}-{\bf{g}})-
{\bf{a}}^2+{\bf{h}}{\bf{a}}-{\bf{b}}^2-\overline{{\bf{b}}}^2 \right]
\end{eqnarray}
where the last factor is just the quadric $\mathrm{Q}_4$ and the first two are trivial extraneous terms.

Considering these results we can now factorize the component $\mathrm{I}_2^{(2)}$ as follows,
\begin{eqnarray}
\mathrm{I}_2^{(2)}&=&{\bf{g}}({\bf{q}}-{\bf{g}})({\bf{h}}-{\bf{a}})^3+
{\bf{a}}({\bf{q}}-{\bf{g}})^2({\bf{h}}-{\bf{a}})^2-
2{\bf{b}}\overline{{\bf{b}}}({\bf{q}}-{\bf{g}})({\bf{h}}-{\bf{a}})^2 \nonumber \\
&=&({\bf{q}}-{\bf{g}})({\bf{h}}-{\bf{a}})^2\left[ {\bf{g}}({\bf{h}}-{\bf{a}})
+{\bf{a}}({\bf{q}}-{\bf{g}})-2\overline{{\bf{b}}}{\bf{b}}\right] \nonumber \\
&=&({\bf{q}}-{\bf{g}})({\bf{h}}-{\bf{a}})^2\left[ {\bf{h}}{\bf{q}}-{\bf{c}}^2
-{\bf{q}}^2\right]
\end{eqnarray}
giving rise to the quadric $\mathrm{Q}_3$.

The final step is to assure the compatibilization of the fractions
$\omega_1({\bf{x}}_2,{\bf{y}}_2)$ and 
$\omega_2({\bf{x}}_2,{\bf{y}}_2)$ with the algebraic curve
$\mathrm{E}_2({\bf{x}}_2,{\bf{y}_2})$. The elimination of the common variables
${\bf{x}}_2,{\bf{y}_2}$ leads us to a single constraint, namely
\begin{equation}
\label{TRAVA}
\left[\omega_1({\bf{x}}_2,{\bf{y}}_2)^2+\omega_2({\bf{x}}_2,{\bf{y}}_2)^2\right]^2-\mathrm{U}\omega_1({\bf{x}}_2,{\bf{y}}_2)\omega_2({\bf{x}}_2,{\bf{y}}_2)^2+2\left[\omega_1({\bf{x}}_2,{\bf{y}}_2)^2-\omega_2({\bf{x}}_2,{\bf{y}}_2)^2\right]+1=0.
\end{equation}

By extracting the functions  
$\omega_1({\bf{x}}_2,{\bf{y}}_2)$ and $\omega_2({\bf{x}}_2,{\bf{y}}_2)$
from the components $\mathrm{I}_2^{(4)}$ and $\mathrm{I}_2^{(5)}$ the 
constraint (\ref{TRAVA}) becomes a polynomial in the
$\mathrm{R}$-matrix weights. This leads to 
the last quadric $\mathrm{Q}_5$ by considering similar simplifications as done
above.

\addcontentsline{toc}{section}{Appendix B}
\section*{\bf Appendix B: Fibration Analysis }
\setcounter{equation}{0}
\renewcommand{\theequation}{B.\arabic{equation}}

In order to study the properties of a generic fiber one can take 
an affine point of $\overline{\mathrm{E}}_2$ such as 
$[c_0,d_0,c_0^2+d_0^2,1]$ where the coordinates
$c_0$ and $d_0$ are constrained by,
\begin{equation}
\label{CONFI}
(c_0^2+d_0^2)^2+\mathrm{U}c_0d_0-1=0.
\end{equation}

The fiber $\pi^{-1}$ is an 
algebraic variety $\subset \mathbb{C}[
{\bf{a}},{\bf{b}},\overline{{\bf{b}}},{\bf{g}}]$ 
described by the following polynomials,
\begin{eqnarray}
\tilde{\mathrm{Q}}_1 &\equiv&  {\bf{b}}\overline{{\bf{b}}}+{\bf{a}}{\bf{g}}-c_0^2=0,
\nonumber \\
\tilde{\mathrm{Q}}_2 &\equiv& {\bf{b}}\overline{{\bf{b}}}+{\bf{g}}({\bf{a}}-1)-
(c_0^2+d_0^2){\bf{a}}
+c_0^2=0, \nonumber \\
\tilde{\mathrm{Q}}_4 &\equiv& {\bf{b}}^2+\overline{{\bf{b}}}^2+{\bf{g}}^2
-(c_0^2+d_0^2){\bf{g}}
+{\bf{a}}({\bf{a}}-1)=0 
\end{eqnarray}

Using the software Singular we found that 
$\pi^{-1}$ turns out to be an irreducible non singular curve of genus one. 
Further information on such elliptic fibration can be obtained by eliminating
the variables $\overline{{\bf{b}}}$ and ${\bf{b}}$ with the help of the quadrics
$\tilde{\mathrm{Q}}_1$ and
$\tilde{\mathrm{Q}}_2$. After using Eq.(\ref{CONFI}) 
the polynomial $\tilde{\mathrm{Q}}_3$ becomes,
\begin{eqnarray}
\mathrm{C}&=&({\bf{a}}^2+{\bf{b}}^2)^2-c_0^4(2{\bf{a}}-1)(2{\bf{a}}^2+2{\bf{b}}^2-2{\bf{a}}+1)
-\mathrm{U}c_0d_0{\bf{a}}\left[{\bf{a}}^3+(1+{\bf{a}}){\bf{b}}^2\right] \nonumber \\
&-&2c_0^2d_0^2\left[(2{\bf{a}}-1){\bf{a}}^2+(2{\bf{a}}+1){\bf{b}}^2\right]
\end{eqnarray}

We end up with a quartic curve on the variables ${\bf{a}}$ and ${\bf{b}}$ which possess
two double points as singularities. These are the simplest singular points we can have
and the curve $\mathrm{C}$ can be desingularized by means of
a single birational transformation bringing it into the Weierstrass form. Let us
denote by $x_0$ and $y_0$ the corresponding affine Weierstrass coordinates then 
the inverse birational map is,
\begin{equation}
x_0=\frac{c_0d_0 \tilde{x}_0}{c_0^2({\bf{a}}-1)^2+(d_0{\bf{a}})^2},~~
y_0=\frac{c_0d_0 \mathrm{U} \tilde{y}_0}{c_0^2({\bf{a}}-1)^2+(d_0{\bf{a}})^2}.
\end{equation}

$\bullet$ The variable $\tilde{x}_0$:
\begin{eqnarray}
\tilde{x}_0&=&\frac{2\alpha_1^2}{\mathrm{U}}{\bf{a}}^2
\left[(d_0^2-5c_0^2){\bf{a}}+\frac{3\IM}{2}(d_0^2-3c_0^2){\bf{b}}\right]
+2\alpha_1{\bf{a}}({\bf{a}}+\IM{\bf{b}})\left[{\bf{b}}^2+\frac{\alpha_1^2}{\mathrm{U}^2}{\bf{a}}^2\right]
+2\alpha_1c_0
(\IM\alpha_5{\bf{b}}+\alpha_6{\bf{a}}){\bf{a}} \nonumber \\
&-&\alpha_2\left[2(2{\bf{a}}-1){\bf{b}}^2+\IM\alpha_3{\bf{b}} +\alpha_4{\bf{a}}\right]
-\IM\mathrm{U}(3c_0^2-d_0^2){\bf{b}}^3+\alpha_7{\bf{c_0}}^3,
\end{eqnarray}
where the coefficients $\alpha_1,\cdots,\alpha_7$ are determined in 
terms of the coordinates $c_0$ and
$d_0$ as follows,
\begin{eqnarray}
&&\alpha_1 = (c_0^2+d_0^2)\mathrm{U},~~
\alpha_2=c_0^2\mathrm{U},~~
\alpha_3=1+4c_0^4-12c_0^2d_0^2-2c_0d_0\mathrm{U}, \nonumber \\
&&\alpha_4=\frac{32c_0d_0}{3\mathrm{U}}+16c_0^4+\frac{11c_0d_0\mathrm{U}}{6}-2,~~
\alpha_5=6c_0^3-6c_0d_0^2-\frac{d_0\mathrm{U}}{2}, \nonumber \\
&&\alpha_6=\frac{8d_0}{3\mathrm{U}}+9c_0^3-3c_0d_0^2-\frac{d_0\mathrm{U}}{24},~~
\alpha_7=\frac{16d_0}{3}+4c_0^3\mathrm{U}-4c_0d_0^2\mathrm{U}-\frac{d_0\mathrm{U}^2}{12}.
\end{eqnarray}

$\bullet$ The variable $\tilde{y}_0$:
\begin{eqnarray}
\tilde{y}_0&=&4c_0d_0\alpha_1{\bf{a}}^2({\bf{b}}-\IM{\bf{a}})({\bf{b}}^2+\frac{\alpha_1^2}{\mathrm{U}^2}{\bf{a}}^2)
+2\frac{\alpha_1^2}{\mathrm{U}^2}{\bf{a}}^3\left[(\beta_1+\alpha_1c_0d_0){\bf{b}}-\IM\beta_1{\bf{a}}\right]
+c_0\frac{\alpha_1}{\mathrm{U}}{\bf{a}}^2(2\IM\beta_2{\bf{a}}-\frac{3}{2}\beta_3{\bf{b}}) \nonumber \\
&+&2{\bf{a}}{\bf{b}}^2\left[(\beta_4+\alpha_1c_0d_0){\bf{b}}-\IM\beta_4{\bf{a}}\right]
+4 \IM c_0^2{\bf{b}}^2\left[(\frac{2\alpha_1\alpha_2}{\mathrm{U}^2}-1)(2{\bf{a}}-1)-\frac{c_0d_0\mathrm{U}}{2}\right]
\nonumber \\
&+&\frac{c_0}{2}(\beta_5{\bf{b}}^3-4\IM\beta_6c_0^2{\bf{a}}^2+\beta_7c_0{\bf{a}}{\bf{b}}-4\IM c_0^2\beta_8{\bf{a}}+\beta_9c_0^2{\bf{b}}+8\IM\beta_{10}c_0^3),
\end{eqnarray}
where the coefficients $\beta_1,\cdots,\beta_{10}$ are given by,
\begin{eqnarray}
&&\beta_1= 2c_0^2 - 2d_0^2 - 13c_0^3d_0\mathrm{U} + 
      3c_0d_0^3\mathrm{U},~~
\beta_2=\frac{\alpha_1}{\mathrm{U}}(8c_0-33c_0^2d_0\mathrm{U}-d_0^3\mathrm{U}) +24c_0d_0^2(2c_0d_0\mathrm{U}-1), \nonumber \\
&&\beta_3=\frac{\alpha_1}{\mathrm{U}}(8c_0-35c_0^2d_0\mathrm{U}-3d_0^3\mathrm{U}) +32c_0d_0^2(2c_0d_0\mathrm{U}-1),~~
\beta_4= 2c_0^2 - 2d_0^2 - 7c_0^3d_0\mathrm{U} + c_0d_0^3\mathrm{U}, \nonumber \\
&&\beta_5 = -8c_0 + 32c_0^3d_0^2 + 32c_0d_0^4 + 
     17c_0^2d_0\mathrm{U} + d_0^3\mathrm{U}, \nonumber \\
&&\beta_6 = 12c_0^3 - 36c_0d_0^2 - d_0\mathrm{U} - 
     40c_0^4d_0\mathrm{U} + 24c_0^2d_0^3\mathrm{U} + 
     c_0d_0^2\mathrm{U}^2, \nonumber \\
&&\beta_7 = 24 - 192c_0^2d_0^2 - 58c_0d_0\mathrm{U} - 
     64c_0^5d_0\mathrm{U} + 192c_0^3d_0^3\mathrm{U} + 
     35c_0^2d_0^2\mathrm{U}^2 - d_0^4\mathrm{U}^2, \nonumber \\
&&\beta_8 = -8c_0^3 + 8c_0d_0^2 + 32c_0^5d_0^2 + 
     32c_0^3d_0^4 + d_0\mathrm{U} + 24c_0^4d_0\mathrm{U} - 
     8c_0^2d_0^3\mathrm{U} - c_0d_0^2\mathrm{U}^2, \nonumber \\
&&\beta_9 = -8c_0^3 - 40c_0d_0^2 + 128c_0^5d_0^2 + 
     128c_0^3d_0^4 - 3d_0\mathrm{U} + 36c_0^4d_0\mathrm{U} + 
     36c_0^2d_0^3\mathrm{U} + 2c_0d_0^2\mathrm{U}^2, \nonumber \\
&&\beta_{10}=\frac{\alpha_1}{\mathrm{U}}(8c_0^2d_0^2-1) +c_0d_0(3c_0^2+d_0^2)\mathrm{U}.
\end{eqnarray}

The corresponding Weierstrass equation for the variables $x_0$ and $y_0$ has been
presented in the main text, see Eq.(\ref{WEIF}). The same analysis can be performed
for the special fiber at the closed set ${\bf{h}}=0$. Once again we find a non-singular
genus one curve which $\mathrm{J}$-invariant is the same as that of the generic fiber given by
Eq.(\ref{JIN3}). This means that the surface $\mathrm{V}$ is normalized in terms
of the product of two elliptic curves.

\end{document}